
\input phyzzx
\hoffset=0.375in
\overfullrule=0pt

\def\dol{{d_{\rm ol}}}
\def\dls{{d_{\rm ls}}}
\def\dos{{d_{\rm os}}}

\def\kms{{\rm km}\,{\rm s}^{-1}}

\def\rot{{\rm rot}}
\twelvepoint
\font\bigfont=cmr17
\centerline{}
\bigskip
\bigskip
\centerline{\bigfont K Band Microlensing of the Inner Galaxy}
\bigskip
\centerline{{\bf Andrew Gould}\foot{Alfred P.\ Sloan Foundation Fellow}}
\smallskip
\centerline{Dept of Astronomy, Ohio State University, Columbus, OH 43210}
\smallskip
\centerline{e-mail gould@payne.mps.ohio-state.edu}
\bigskip
\centerline{\bf Abstract}
\singlespace

Microlensing searches toward the inner galaxy $(|l|,|b|\leq 22.\hskip-2pt'5)$
have several major advantages.  First, the event rate is strongly dominated
by bulge-bulge lensing events where both the source and lens lie in the bulge.
Second, these bulge-bulge events have very short time scales $t_e\sim 2\,$days
and are therefore easily distinguished from the less frequent and much longer
bulge-disk and disk-disk events.  Third, since the optical depth is similar to
that at higher impact parameters, while the events are shorter, the event rate
is high $\Gamma\sim 3\times 10^{-7}\rm day^{-1}$.  Fourth, because the
Einstein rings are small, $r_e\sim 5\times 10^{12}\,$cm, and the source stars
are large $r_s\gsim 10^{12}\,$cm, the lens will transit the face of the source
for a significant fraction $(\sim 20\%)$ of events.  For these transit events
it will often be possible to measure a second lens parameter, the angular
Einstein radius (or proper motion).  In addition to the bulge-bulge events,
the optical depth of the disk is $\sim 7$ times larger toward the inner Galaxy
than toward Baade's Window.  A microlensing search toward the inner Galaxy can
be carried out by making frequent $(\sim4\,\rm day^{-1})$ K band images of a
large area $\sim 0.5\,\rm deg^2$ to a depth of $K\sim 16$, and hence requires
either a $1024^2$ infrared array on a dedicated 2m telescope or four such
arrays on a 1m telescope.

\bigskip
Subject Headings:  dark matter -- Galaxy: stellar content --
gravitational lensing -- stars: low-mass
\vskip 0.2in
\centerline{submitted to {\it The Astrophysical Journal Letters}:
December 20, 1994}
\smallskip
\centerline{Preprint: OSU-TA-27/94}

\endpage
\normalspace
\chapter{Introduction}

	Microlensing searches toward the inner Galaxy in K band could
help constrain the mass spectrum of the lensing objects now being
detected at a rapid rate in ongoing microlensing searches.  Lensing
searches toward the inner Galaxy $\lsim 30'$ from the Galactic center
possess several unique features.  Most notably, the events due to bulge
lenses have very short $\sim 2\,$day time scales, so they can be clearly
distinguished from the much longer events due to disk lenses.  Hence,
an inner-Galaxy microlensing search can provide information that can
clarify the major puzzles arising from current observations along lines of
sight at $\sim 4^\circ$ from the Galactic center.
Because of high extinction, this
search can be carried out only in the K band.

	Optical microlensing searches toward the Galactic bulge
by Alcock et al.\ (1994 MACHO) and Udalski et al.\ (1994 OGLE)
have detected more than 50 candidate events to date.  The event rate is
substantially greater than was anticipated by Paczy\'nski (1991)
and Griest et al.\ (1991) when they originally proposed the experiment
as a method to probe the mass distribution of the Galactic disk.
Bennett et al.\ (1994) estimate the mean optical depth toward fields near
Baade's Window to be $\tau\sim 3\times 10^{-6}$, more than five times higher
than that expected from the disk.
Kiraga \& Paczy\'nski (1994) pointed out that at least some of this
shortfall could be accounted for by the `standard' Kent (1992)
model of the Galactic bulge with a mass
$M_{bul}\sim 1.2\times 10^{10}\,M_\odot$,
which yields an optical depth $\tau_{bul}\sim 0.7\times 10^{-6}$ in
addition to that due to the disk.  Paczy\'nski et al.\ (1994) suggested
that a triaxial Galactic bulge could substantially enhance the lensing rate.
Han \& Gould (1994) used the virial theorem to estimate the mass of an
elongated bulge as modeled by Dwek et al.\ (1994) on the basis of data
from the {\it Cosmic Background Explorer (COBE)}.  They found $M_{bul}\sim
1.8\times 10^{10} M_\odot$ and $\tau_{bul}\sim 1.3\times 10^{-6}$.
Zhao, Spergel, \& Rich (1995) obtained a similar result from a more detailed
model.  Blum (1994) pointed out that Han \& Gould had neglected an important
term related to figure rotation in the virial equation. He derived a further
upward revision to as much as
$M_{bul}\sim 3.2\times 10^{10}M_\odot$ with corresponding
optical depth $\tau_{bul}\sim 2.3\times 10^{-6}$.  Note that Blum's estimate
would imply that $\sim 38\%$ of the mass of the Galaxy interior to the solar
orbit is in the bulge, and requires a mass-to-light ratio for the bulge of
$M/L_V\sim 16$ (see Binney \& Tremaine 1987).

	These rapidly escalating estimates of the bulge mass and
mass-to-light ratio raise the
question of what the bulge is made of.  In particular, one would like to know
the spectrum of lens masses, $M$.  The lensing events observed to date
are of limited help in resolving this question for two reasons.  First,
the only information available about the events is their time scale,
$t_e$.  While this quantity generally scales $\propto M^{1/2}$,
the mass associated with individual events can be estimated only to within
an order of magnitude (Griest et al.\ 1991).  Second, and greatly complicating
the first problem, it is not known whether any individual lens that is
detected is in the bulge or in the Galactic disk.  The two classes of
events have very similar time scales (Han \& Gould 1994).  Furthermore,
as pointed out by Bennett et al.\ (1994), the optical depth appears to
rise rapidly as one approaches the Galactic plane, perhaps indicating that
a considerable fraction of all events are due to disk lenses.  One would
like to be able to examine a clean sample of bulge lenses.

	Here I show that bulge lenses in the inner Galaxy present such
a clean sample.  Unlike the lenses found at higher impact parameter,
these objects give rise to very short events that can easily be distinguished
from foreground disk lenses.  In particular, the low-mass lenses which give
rise to the shortest events are especially easy to recognize.  Moreover,
for a large fraction $(\sim 20\%)$ of these inner-Galaxy events,
the lens will transit the source, making it possible to measure the
Einstein Ring radius.  These measurements can help further constrain
the mass spectrum.
A K band
lensing search of the inner Galaxy can therefore yield important clues about
the mass spectrum of the bulge which are not available from data taken
along other lines of sight.

\chapter{Microlensing By An Isothermal Sphere}

	In order to illustrate the fundamental features of microlensing
toward the inner Galaxy, I analyze a simplified model.  I assume that the
lenses are distributed in a singular isothermal sphere with
$$\rho(r) = {v_\rot^2\over 4 \pi G r^2},\eqn\rhoiso$$
where $v_\rot$ is the circular rotation speed which characterizes the density
distribution.  The Einstein ring radius is given by
$r_e = (4 G M \dol\dls/ c^2 \dos)^{1/2}$
where $\dol$, $\dls$, and $\dos$ are the distances between the observer,
source, and lens.  I assume that the lens lies close
to the source, $\dls\ll\dos$, so that
$$r_e=\sqrt{4 G M \dls\over c^2}.\eqn\redef$$
Finally I assume that the source lies exactly at the distance of the Galactic
center, $\dos=R_0$.  This last assumption introduces some distortion into
the final result which I discuss below.  However, it also makes the
problem analytically tractable and so greatly facilitates the discussion.

	Let $t_e= r_e/v$ be the characteristic time of the event,
where $v$ is the (two-dimensional) transverse speed of the lens relative to
the observer-source line of sight.  If both the sources and lenses have
Gaussian velocity distributions
$f(u_x,u_y) = (\pi v_\rot^2)^{-1}\exp[-(u_x^2+u_y^2)/v_\rot^2]$,
then the relative
speed distribution is
$$f(v)d v = {v \over v_\rot^{2}}\exp\biggl(-{v^2\over 2v_\rot^2}\biggr)
d v.\eqn\speeddis$$
The cross section to lensing is $2 r_e$, so that the
event rate for fixed lens mass $M$ and for event times $t_e'<t_e$ is
$$\Gamma(t_e) = \int_0^\infty d\dls\,2 r_e(\dls,M){\rho(\dls)\over M}
\int_{r_e/t_e}^\infty d v\, v f(v)
.\eqn\dGdo$$
Hence, the differential rate is
$${d\Gamma\over d t_e} = {4\over \pi}\,{v_\rot^2\over c^2}t_e^{-2}\int_0^\infty
d z {z^2\over z^2 + (t_b/t_e)^4}\exp(-z),
\eqn\dgdom$$
where $z$ is a dummy variable,
$$t_b\equiv {(2 G M b)^{1/2}\over v_\rot c}= 1.7\,{\rm day}\,
\biggl({M\over 0.2 M_\odot}\biggr)^{1/2}\,\biggl({b\over 50\,\rm pc}
\biggr)^{1/2}\,
\biggl({v_\rot\over 200\,\kms}\biggr)^{-1},\eqn\tbeq$$
and $b$ is the impact parameter to the Galactic
center.
The total event rate is
$$\Gamma = \sqrt{\pi\over 2}{v_\rot^2\over c^2}t_b^{-1} =
3\times 10^{-7}\,{\rm day}^{-1}\,
\biggl({M\over 0.2 M_\odot}\biggr)^{-1/2}\,\biggl({b\over 50\,\rm pc}
\biggr)^{-1/2}\,
\biggl({v_\rot\over 200\,\kms}\biggr)^{3}.\eqn\gamtot$$

	In the two limits, the rate is
$${d\Gamma\over d t_e} = {8\over \pi}{v_\rot^2\over c^2}{t_e^{2}\over t_b^4}
\qquad
(t_e\ll t_b),\eqn\gamlow$$
and
$${d\Gamma\over d t_e} = {4\over \pi}{v_\rot^2\over c^2}t_e^{-2}\qquad
(t_e\gg t_b).\eqn\gamhigh$$

	The optical depth associated with these lensing rates is formally
infinite because the range of the $\dls$ integration has been extended
to infinity.  In practice, this range is cut off at some radius $R_c$
which is the shorter of the physical extent of the isothermal bulge or
the distance $\dls\sim R_0/2$ where equation \redef\ breaks down.  The
optical depth is then $\tau\sim (v_\rot^2/c^2)
\ln(R_c/b)\sim 1.5\times 10^{-6}$.

\chapter{Discussion}

	The above calculation shows that lensing events observed toward the
inner Galaxy have two notable features:
they are much shorter and more plentiful
than bulge-bulge events observed toward optical windows at high impact
parameter such as Baade's Window ($b\sim 500\,$pc).  The fact that the
events are short makes them easy to distinguish from lensing of bulge
or disk stars by disk lenses (disk-bulge and disk-disk events respectively).
By contrast, the time scales of bulge-bulge and disk-bulge events seen
toward Baade's Window are very similar, making it impossible to distinguish
between the two by measuring $t_e$ (Han \& Gould 1994).  Observations toward
Baade's Window and
similar fields show a noticeable absence of short events $t_e<6\,$days,
implying in particular that disk-bulge events must typically be longer than
this limit.  Hence, events observed toward the inner Galaxy with time
scales $\sim 2\,$days could be catalogued as bulge-bulge events with high
confidence.  Thus, the mass spectrum of bulge objects could be estimated
without worrying about disk contamination.

	The high event rate means that good statistics could be obtained
over a few bulge seasons, in spite of the fact that it would be possible
to monitor only $\sim 10^6$ stars in the inner Galaxy (see \S\ 4).  During
a six-month season, one could expect to observe $\sim 50$ events.

	The event rate from disk-bulge events would be lower than the
bulge-bulge rate, but would still be very significant.  At present, there
is considerable debate over what fraction of the events seen toward Baade's
Window and similar fields is due to disk lenses.  Paczy\'nski (1991) and
Griest et al.\ (1991) originally predicted an optical depth in this directions
of $\sim 5\times 10^{-7}$.  However, Alcock et al.\ (1994) and Bennett et al.\
(1994) have speculated that the disk may be more massive than the standard
`no missing mass' disk on which these predictions were based.  Regardless of
the exact normalization or the functional form of the disk, one can be certain
that the optical depth due to disk objects is higher toward the Galactic
center than toward Baade's Window.  For example,  for an exponential disk
with scale height $h\sim 325\,$pc and scale length $H\sim 3000\,$pc, the
optical depth is a factor $\sim 7$ higher.
See e.g., Griest et al.\ (1991).
Thus, the inner-Galaxy fields would provide
a rich source of information about disk microlensing.  Since these events
would typically be much longer than the bulge-bulge events, the disk-bulge
events could generally be recognized as such.

	Finally, for a significant fraction of the bulge-bulge events,
the lens would transit the source, making it possible to measure the
size of the angular Einstein ring, $\theta_e=r_e/\dol$
(Gould 1994; Nemiroff \& Wickramasinghe 1994).
The cross section for such transit events is $2 r_s$ where $r_s$ is the
radius of the source star.  Hence, the rate for an ensemble of source stars is
$$\Gamma_{\rm tran} = 2\VEV{v}\VEV{r_s}\int_0^\infty d\dls{\rho(\dls,M)\over M}
=\sqrt{2\pi}{v_\rot^3\over 8 G M b}\VEV{r_s},\eqn\gtran$$
where $\VEV{r_s}$ is the mean radius of the source stars.
Comparison with equation \gamtot\ shows that a fraction,
$${\Gamma_{\rm tran}\over \Gamma} = {\VEV{r_s}c\over (8 G M b)^{1/2}},
\eqn\gamrat$$
of the events would be transits.  For typical numbers, $M=0.2\,M_\odot$,
$b=50\,$pc, $\VEV{r_s}=22\,R_\odot$ (see \S\ 4), this fractions is $\sim 26\%$.
Because $\dol\sim \dos$, measurement of $\theta_e$ yields the product $M \dls$.
This information in conjunction with the measurement of $t_e$ can help
constrain the mass spectrum.

	The entire analysis given here has assumed that the sources lie exactly
at the midpoint of the lens density distribution.  Realistically, one expects
that the sources will be distributed as the lenses. I have elsewhere shown that
this approximation leads to an underestimate of the optical depth
by a factor $\alpha=\int_0^\infty d z [1 - F(z)]/\int_0^\infty
d z [1 - F^2(z)]$ where $F(z)$ is the cumulative distribution of sources as
a function of distance $z$ from the midpoint of the distribution (Gould 1995).
The size of this effect ranges from $2/3\lsim\alpha\lsim 3/4$. It arises
because the characteristic distances $\dls$ are typically underestimated
by a factor $\alpha$.  For this reason, the effect of including a realistic
distribution is to increase the typical event times by $\alpha^{-1/2}$ and
to increase the event rate by a similar amount.  Hence, for present purposes
one should simply scale all rates and time scales given in equations
\dgdom-\gamhigh\ upward by $\sim 1.2$.  Similarly, equation \gamrat\ should
be reduced by a factor $\sim 1.2$.

\chapter{Practical Requirements}

	The inner Galaxy is heavily obscured, with visual extinction
$A_V\sim 20$--30 in most areas and much higher values in isolated spots.  Thus
K band photometry is essential to any microlensing search of this region.
To make my estimates of the requirements of such a search, I assume
$A_V=30$ over 80\% of the region and $A_V=\infty$ over the rest.  I assume
$1.\hskip-2pt '' 5$ seeing and scale from the experience of the MACHO
group (D.\ Bennett 1994, private communication) that in their crowding-limited
fields they are able to monitor $10^6$ stars $\rm deg^{-2}$ in $2''$ seeing
with $0.\hskip-2pt '' 6$ pixels.  I infer that in crowding-limited K band
images, one could resolve $1.1\times 10^6$ stars in a $45'$ square field
imaged with $0.\hskip-2pt '' 45$ pixels.  Allowing for total obscuration
over 20\% of the field, this is still $9\times 10^5$ stars.  The area could
be covered in 36 exposures of a $1024^2$ infrared array.

	To find the crowding limit, I normalize the K band luminosity function
from Baade's Window (G.\ Tiede and J.\ Frogel 1994, private communication)
to the bright K-band counts of an inner-Galaxy field with measured
extinction $A_K=2$ (J.\ Frogel 1994, private communication) and find that
there are $\sim 2\times 10^6\,\rm deg^{-2}$ stars to $K_0\sim 13$.  Hence,
the exposures must reach $K=16$ to achieve the crowding limit at the assumed
$A_K=3$ and $1.\hskip-2pt ''5$ seeing.

	For average conditions of a Chilean winter $(K_{\rm sky}\sim 13$
mag arcsec${}^{-2}$), a K=19.5 source contained within 1 arcsec${}^2$ has
signal to noise 10 in 35 minutes of exposure on a 1.8 m telescope (J.\ Frogel
1994, private communication).  Scaling from this experience, I estimate
that an isolated $K=16$ star can be photometered with accuracy $\sim 3.5\%$
with a 10 minute exposure on a 1 m telescope.  This is similar to the
typical signal to noise achieved by MACHO at the crowding limit in their
fields in the Large Magellanic Clouds (LMC).  I conclude that 10 minute
exposures
would suffice to obtain the $\sim 10\%$ photometry typically
achieved by MACHO in their crowded LMC fields.

	A mosaic of 4 $1024^2$ arrays mounted on a 1 m telescope could
therefore cover the inner bulge in $\sim 2\,$hrs, allowing for reasonably
fast read out and pointing.  Hence, stars could be monitored $\sim 4$ times
per clear winter night.  In the spring and fall, only 2 or 3 observations
would be possible.  This would be adequate to recognize and measure the
short $\sim 2\,$day events which are expected to be typical.  Alternatively,
one could achieve the same signal to noise on a 2 m telescope in only 150 s,
implying that similar results could be obtained on a 2 m with a single
$1024^2$ array.

	I now justify the estimate for the mean radius of a source star
$\VEV{r_s}\sim 22 R_\odot$ made near the end of the last section to calculate
the fraction of transit events.  In the Raleigh-Jeans limit, the radius
of a star scales $r_s\propto T^{-1/2}10^{-M_K/5}$, i.e., it is only
weakly dependent on temperature, $T$.  Adopting this framework, I estimate
the stellar radius to be $r_s = 11\times 10^{(13-K_0)/5} R_\odot$, where I have
assumed that the source stars are a factor 0.7 cooler than the Sun,
$M_{K,\odot}\sim 3.3$, and $R_0=8\,$kpc.  Then
integrating over the K band luminosity function, I find
$\VEV{r_s}=22\,R_\odot$.  Since, the Raleigh-Jeans limit does not strictly
apply and since the brightest stars are actually cooler than I have
assumed, the true value is somewhat higher.

{\bf Acknowledgements}:  I would like to thank J.\ Frogel and G.\ Newsom for
making several helpful suggestions and comments.

\endpage
\Ref\Alcock{Alcock, C., et al.\ 1994, ApJ, in press}
\Ref\ben{Bennett, D.\ et al.\ 1994, preprint}
\Ref\blum{Blum, R.\ 1994, ApJ Letters, submitted}
\Ref\bt{Binney, J.\ \& Tremaine, S.\ 1987, Galactic Dynamics (Princeton:
Princeton Univ.\ Press)}
\Ref\dwek{Dwek, E.\ 1994, ApJ, submitted}
\Ref\gtwo{Gould, A.\ 1994, ApJ, 421, L71}
\Ref\gtwo{Gould, A.\ 1995, ApJ, 441, 000}
\Ref\grie{Griest, K.\ et al.\ 1991, ApJ, 372, L79}
\Ref\gtwo{Han, C.\ \& Gould, A.\ 1994, ApJ, submitted}
\Ref\ken{Kent, S.\ M.\ 1992, ApJ, 387, 181}
\Ref\kir{Kiraga, M.\ \& Paczy\'nski, B.\ 1994, ApJ, 430, 101}
\Ref\nem{Nemiroff, R.\ J.\ \& Wickramasinghe, W.\ A.\ D.\ T.\ ApJ, 424, L21}
\Ref\Pa{Paczy\'nski, B., Stanek, K.\ Z., Udalski, A., Szyma\'naski, M.,
Kaluzny, J, Kubiak, M., \& Szyma\'nski, W.\
 1994, ApJ, 435, L63}
\Ref\Pac{Paczy\'nski, B.\ 1991, ApJ, 371, L63}
\Ref\ogle{Udalski, A., Szyma\'nski, J.,  Stanek, K.\ Z.,
Ka{\l}u\.zny, J., Kubiak, M.,
Mateo, M., Krzemi\'nski W., Paczy\'nski, B., \& Venkat, R.\ 1994,
Acta Astronomica, 44, 165}
\Ref\Zhou{Zhao, H.\ S., Spergel, D.\ N., \& Rich, R.\ M.\ 1995 ApJ, in press}
\refout
\endpage
\endpage
\end